\documentclass[aps,preprint]{revtex4}%
\usepackage{amsfonts}
\usepackage{amsmath}
\usepackage{amssymb}
\usepackage{graphicx}%
\begin{document}
\title{Relativity in Introductory Physics}
\author{William E. Baylis}
\affiliation{Department of Physics, University of Windsor, Windsor, ON, Canada N9B 3P4}
\pacs{PACS number}

\begin{abstract}
A century after its formulation by Einstein, it is time to incorporate special
relativity early in the physics curriculum. The approach advocated here
employs a simple algebraic extension of vector formalism that generates
Minkowski spacetime, displays covariant symmetries, and enables calculations
of boosts and spatial rotations without matrices or tensors. The approach is
part of a comprehensive geometric algebra with applications in many areas of
physics, but only an intuitive subset is needed at the introductory level. The
approach and some of its extensions are given here and illustrated with
insights into the geometry of spacetime.

\end{abstract}
\date[Date text]{date}
\received[Received text]{date}

\accepted[Accepted text]{date}

\maketitle

\section{Introduction}

Special relativity, which has now been with us for about one century, presents
a new world view or \emph{paradigm}\cite{Kuhn70}\emph{ }of physics. It revises
the concepts of time and space from those assumed in Newtonian and Galilean
physics, with consequent changes, for example, in what we mean by simultaneity
and how we \textquotedblleft add\textquotedblright\ velocities. It also
establishes the framework for fundamental symmetries of electromagnetic
phenomena and much of modern physics. Such symmetries provide new approaches
to many problems, simplifying many computations, and they are frequently
important even at low (\textquotedblleft nonrelativistic\textquotedblright)
velocities, as demonstrated below. Yet, it appears that the physics community
has still not completed the paradigm shift: relativity is commonly taught as a
complicating correction to Newtonian mechanics, and if elementary
electromagnetic theory texts mention it at all, it is usually only later in
the text after the basic laws and phenomena have been discussed. Is it not
time to integrate relativity more tightly into the early physics curriculum?

One reason for delaying the introduction of relativity in the curriculum can
be traced to the conceptual inertia of the educational process: those teaching
today learned Newtonian mechanics before relativity and typically feel that a
good grounding in Galilean transformations is needed before Lorentz
transformations can be understood. It may be argued that the remnants of
Aristotelian notions must be cleared away before students are faced with
concepts such as the observer-dependence of space and time. However, the
relativistic approach is the correct one, and in some respects it may be more
intuitive to the beginner than Newtonian mechanics. It is doubtful that it
serves the students' best interests to ingrain faulty concepts such as
universal time and instantaneous interactions at a distance.

An important practical reason for delaying the introduction of relativity is
mathematical. Introductory treatments of relativity that go beyond basic
concepts to practical calculations almost all use matrices or tensors at the
expense of the vector notation common in Newtonian physics. Matrix and tensor
elements are much less effective than vectors at conveying the geometry that
is so critical for our physical understanding. (Some efforts to use the
geometrical power of differential forms has been made, but they face a
conceptual barrier in the abstraction of vector concepts and have been largely
reserved to treatments of general relativity.)

This paper advocates an alternative treatment: an algebraic approach based on
a simple extension of vectors in physical space. By replacing the dot and
cross products of vectors with a simpler but more general associative product,
one is led to add scalar time components to vectors. Such objects form a
four-dimensional linear space with Minkowski spacetime metric. The algebra
that results allows relativistic calculations while avoiding matrices and
tensors (although these can be readily derived in the approach). The algebraic
approach advocated here for introductory courses is a subset of the powerful
and well-developed covariant formalism of Clifford's geometric algebra of
physical space (APS).\cite{Bay99,Bay03,Hes99} APS is the Clifford or geometric
algebra of three-dimensional Euclidean space, sometimes denoted $C\!\ell_{3}$
or $\mathcal{G}_{3}.$ It is isomorphic (equivalent in structure) to complex
quaternions, the algebra of Pauli spin matrices, and the even subalgebra of
Hestenes' spacetime algebra (STA).\cite{Hes66, Hes03b}

The following Section emphasizes the importance of relativistic symmetries
even at low velocities. In Section III, we introduce the basic algebra needed
to compute Lorentz transformations and give several examples appropriate for
an introductory course. A discussion of the new geometrical concept of a
bivector and its advantages over the more traditional cross product is given
in Section IV. Extensions to rotor and spinor representations for use in later
courses are discussed in Section V. APS is briefly compared to an alternative
algebraic approach using STA in Section VI, and conclusions are summarized in
the final Section. While the paper is largely pedagogical, several new results
that flow from the APS\ approach are presented. These include a simple
formulation in Subsection III-C of relations between the Minkowskian geometry
of spacetime vectors and the lengths and angles actually measured (in a
Euclidean sense) on a spacetime diagram, a derivation of reduced algebraic
forms for Lorentz rotations in arbitrary spacetime planes Subsection V-B, and
the demonstration in Subsection V-C that any boost of an electromagnetic plane
wave is equivalent to a spatial rotation and dilation.

\section{Relativity at Low Velocities}

\begin{figure}
[tbh]
\begin{center}
\includegraphics[
height=1.7045in,
width=3.0995in
]%
{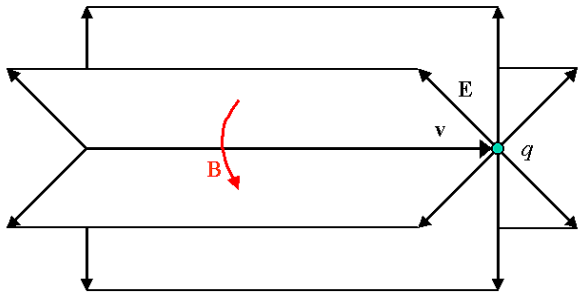}%
\caption{The electric field lines of a moving charge $q$ sweep out spatial
planes that represent the magnetic field $\mathbf{B~.}$}%
\label{FofmovingQ}%
\end{center}
\end{figure}

The importance of relativity is well appreciated for understanding Einstein's
mass-energy relation or for correctly computing such values as threshold
energies for particle production in high-velocity collisions. Less appreciated
is the power of relativistic symmetries at low (\textquotedblleft
nonrelativistic\textquotedblright) velocities, particularly in electromagnetic
theory. One fruitful example is the understanding that the electric and
magnetic fields are aspects of a single covariant electromagnetic field. For
example, the electric field lines of a moving charge sweep out spatial planes
that represent its magnetic field. The magnetic field is the vector normal
(dual) to the planes (see Fig. \ref{FofmovingQ}), as discussed more fully in
Subsection IV.B. The magnetic field that necessarily accompanies the
oscillating electric field in an electromagnetic plane wave has a similar
origin. Indeed, the geometric interpretation of the magnetic field as a
spatial plane and its connection to the electric field can help to demystify
many relations involving $\mathbf{B.}$%

There are also many cases in which a relativistic transformation can simplify
a computation and clarify its significance. The following examples will serve
to illustrate that the importance of relativity is not restricted to exotic
phenomena that occur only with particle velocities close to the speed of light.

\begin{figure}
[tbh]
\begin{center}
\includegraphics[
height=2.6515in,
width=3.282in
]%
{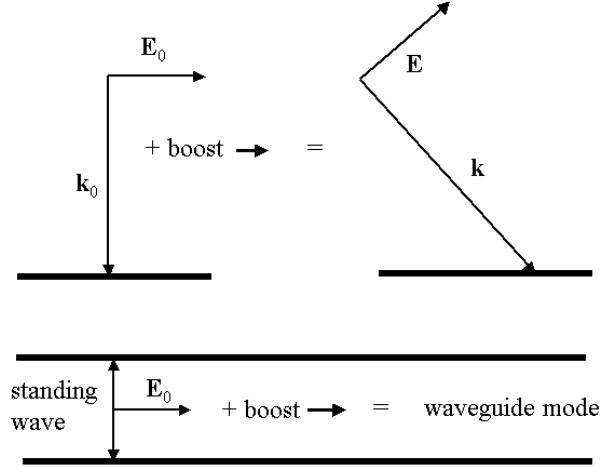}%
\caption{When propagating plane waves are boosted, they are simply rotated and
dilated. As a result, waves obliquely incident on a conductor can be obtained
by boosting normally incident ones. Similarly, waveguide modes can be found by
boosting standing waves.}%
\label{relwo}%
\end{center}
\end{figure}

To calculate the effect of a plane wave incident at an oblique angle on the
plane surface of a good conductor, we can compute the simpler case of normal
incidence and then apply a boost (velocity transformation) in the plane of the
conductor (see Fig. \ref{relwo}). This can be extended to the calculation of
wave-guide modes by boosting standing waves.\cite{Bay99}%

We return to these examples at the end of this paper and show that the boost
of a propagating plane wave and of its wave vector in spacetime is equivalent
to a rotation and dilation (rescaling).

\section{The Algebraic Approach}

The key step in the algebraic method is to introduce an associative product of
vectors. This product, called a \emph{geometric product}, is defined by the
axiom that any vector $\mathbf{p}$ times itself is the scalar product
$\mathbf{p\cdot p,}$ that is, the square length of $\mathbf{p}$ :%
\begin{equation}
\mathbf{p}^{2}=\mathbf{pp}=\mathbf{p\cdot p~.} \label{axiom}%
\end{equation}
The square length of a vector $\mathbf{p}=p_{x}\mathbf{e}_{1}+p_{y}%
\mathbf{e}_{2}+p_{z}\mathbf{e}_{3}$ in (flat, Euclidean) physical space is
given by the orthonormality of the basis vectors $\mathbf{e}_{j}$ to be the
Pythagorean value $\mathbf{p}^{2}=p_{x}^{2}+p_{y}^{2}+p_{z}^{2}~.$
Multiplication of a vector $\mathbf{p}$ by a (real) scalar $\alpha$ scales its
length, and the two vectors $\mathbf{p}$ and $\alpha\mathbf{p}$ are said to be
\emph{aligned.} It follows that the geometric product of any two aligned
vectors is their scalar product. The product of nonaligned vectors will be
considered below.

The product of elements in APS behaves like the product of square matrices.
There are in fact infinitely many matrix representations of APS, but explicit
matrices are never needed. The existence of a single matrix representation
proves that APS exists as a consistent algebra, and once its existence has
been accepted, only the algebra itself is important.

Those familiar with matrices but uncomfortable with vector algebras may find
it initially reassuring to represent the basis vectors $\mathbf{e}_{j}$ by the
three $2\times2$ Pauli spin matrices $\sigma_{j}.$ The vector-space properties
of physical space, namely vector addition, scalar products of vectors, and
multiplication of vectors by scalars, then work just as well as when the
$\mathbf{e}_{j}$ are thought of as three-dimensional column or row vectors, as
is common in elementary vector treatments. However, the alternative
representation of vectors as $2\times2$ matrices also leads naturally to (i)
an associative multiplication of vectors, and (ii) the addition of a fourth
dimension proportional to the unit matrix $\sigma_{0}=\left(
\begin{array}
[c]{cc}%
1 & 0\\
0 & 1
\end{array}
\right)  $, which is linearly independent of the $\sigma_{j}.$ A $2\times2$
representation of the full APS is generated from products of the matrix
representations of vectors. Every element is then some $2\times2$ matrix, and
scalars in particular are represented by the scalar value times $\sigma_{0}%
$.\cite{Bay80} Generally, however, matrices can and probably should be avoided
entirely when introducing the algebra at the first- or second-year university level.

The geometric product is the only product needed in the algebra. As we will
see below, both the dot and cross product of vectors can be expressed in terms
of it. Like the cross product, the geometric product of nonaligned vectors
does not commute. This is seen by setting $\mathbf{p=q+r}$ in the axiom
(\ref{axiom}) to get%
\[
\left(  \mathbf{q+r}\right)  \left(  \mathbf{q+r}\right)  =\mathbf{q}%
^{2}+\mathbf{rq}+\mathbf{qr+r}^{2}=\mathbf{q\cdot q}+2\mathbf{r\cdot
q}+\mathbf{r\cdot r~,}%
\]
which with the help of the axiom gives an expression for the dot product:%
\begin{equation}
\mathbf{r\cdot q=}\frac{1}{2}\left(  \mathbf{rq}+\mathbf{qr}\right)  .
\label{innerprod}%
\end{equation}
If $\mathbf{r}$ and $\mathbf{q}$ are perpendicular, this expression vanishes.
Thus, while aligned vectors commute, perpendicular vectors anticommute. Unlike
the dot and cross products, the geometric product is associative and
invertible: if $\mathbf{p,q,r}$ are any vectors,%
\[
\left(  \mathbf{pq}\right)  \mathbf{r}=\mathbf{p}\left(  \mathbf{qr}\right)
\equiv\mathbf{pqr}%
\]
and%
\[
\mathbf{p}^{-1}=\frac{\mathbf{p}}{\mathbf{p}^{2}}~.
\]
The additional effort required to teach the geometric product is more than
compensated for by avoiding the confusion caused by the non-associative cross
product. Furthermore, unlike the cross product, the geometric product is
easily extended to spaces of more than three dimensions.

\subsection{Spacetime Vectors}

A point $r$ in spacetime depends on the time $t$ as well as on the spatial
position $\mathbf{r}$ in physical space. Now elements of APS may be sums of
vectors and their products, and as we have seen, those products include real
scalars. We might therefore try to represent $r$ in APS by the sum
$r=ct+\mathbf{r~.}$ (The factor of the speed of light $c$ is included to
ensure that all components of $r$ have the same dimensions.) In Clifford
algebras, the sum of a scalar and a vector is commonly called a
\emph{paravector},\cite{Maks89,Porteous95,AblamLoun95,Bay96,Loun97}. We adopt
this name here to distinguish it from a vector in three-dimensional physical
space. A displacement of $r$ is then a spacetime vector represented by%
\begin{equation}
dr=cdt+d\mathbf{r}.\nonumber
\end{equation}
Other spacetime vectors are similarly represented, for example the spacetime
momentum of a particle is a paravector%
\[
p=p^{0}+\mathbf{p~,}%
\]
where $E=p^{0}c$ is the energy and $\mathbf{p}$ the spatial momentum. The sum
is analogous to the sum of a real number and an imaginary number to form a
complex number.

Whether or not paravectors in APS can represent spacetime vectors depends on
how the square length of a paravector is determined. The square length of a
vector $\mathbf{p}$ is simply $\mathbf{p}^{2},$ but the square of a paravector
$p$ is generally not a scalar. The analogy to complex numbers suggests that we
need to multiply $p$ by a conjugate, and what we need is called the
\emph{Clifford conjugate} (or bar conjugate) $\bar{p}=p^{0}-\mathbf{p~,}$
which changes the sign of the vector part. The product%
\begin{equation}
p\bar{p}=\left(  p^{0}+\mathbf{p}\right)  \left(  p^{0}-\mathbf{p}\right)
=\left(  p^{0}\right)  ^{2}-\mathbf{p}^{2}=\bar{p}p
\end{equation}
is always a scalar and can be taken as the \textquotedblleft square
length\textquotedblright\ of $p.$ The minus sign, which appears naturally
here, dictates that paravector space has the geometry of Minkowski spacetime.
It marks an important departure from the Euclidean space of $\mathbf{p}$ since
the \textquotedblleft square length\textquotedblright\ of $p$ and $\bar{p}$
can be either positive, negative, or zero. By replacing $p$ by $q+r,$ the
scalar product of two distinct paravectors $q,r$ is found. It is the scalar
part of the geometric product $q\bar{r}:$
\begin{equation}
\left\langle q\bar{r}\right\rangle _{S}\equiv\frac{1}{2}\left(  q\bar{r}%
+r\bar{q}\right)  =q^{0}r^{0}-\mathbf{q\cdot r}=\left\langle r\bar
{q}\right\rangle _{S}\mathbf{~.}%
\end{equation}
Note that $\overline{q\bar{r}}=r\bar{q}.$ Paravectors $q$ and $r$ are said to
be \emph{orthogonal} if $\left\langle q\bar{r}\right\rangle _{S}=0.$ As long
as $p\bar{p}\neq0,$ $p$ has an inverse%
\begin{equation}
p^{-1}=\bar{p}/\left(  p\bar{p}\right)  .
\end{equation}
This is similar to the inverse of a complex number, but there is now a new
possibility: if $p\bar{p}=0$ but $p\neq0,$ then $p$ is \emph{null} and has no
inverse. Null elements are orthogonal to themselves. They arise for travel at
the speed of light.

The paravector basis is four-dimensional, and we define an orthonormal set
$\left\{  e_{0},e_{1},e_{2},e_{3}\right\}  $ comprising one unit paravector
along the time direction $e_{0}\equiv1$ plus the three orthogonal paravectors
$e_{j}\equiv\mathbf{e}_{j}$ along the spatial axes. Their scalar products give
elements of what is known as the \emph{Minkowski spacetime metric tensor}%
\begin{equation}
\left\langle e_{\mu}\bar{e}_{\nu}\right\rangle _{S}=\left\{
\begin{array}
[c]{rc}%
1, & \mu=\nu=0\\
-1, & \mu=\nu=1,2,3\\
0, & \mu\neq\nu
\end{array}
\right.  .
\end{equation}

\subsection{Lorentz Transformations}

An important paradigm of relativity is that space and time are not absolute
but are mixed by physical \emph{Lorentz transformations,} which may be viewed
as rotations in spacetime. The square length of a spacetime vector is
invariant under such rotations. If $dr=cdt+d\mathbf{r}$ is the displacement of
a particle, the Lorentz-invariant square length of the displacement is%
\begin{equation}
dr~d\bar{r}=c^{2}dt^{2}-d\mathbf{r}^{2}=c^{2}d\tau^{2}.
\end{equation}
It suggests the definition of a Lorentz-invariant proper time $\tau$ as the
time in the commoving inertial frame of the particle, where $d\mathbf{r}=0.$
The dimensionless proper velocity is defined as%
\begin{align}
u  &  =\frac{dr}{cd\tau}=\frac{dt}{d\tau}\left(  1+\frac{d\mathbf{r}}%
{cdt}\right) \nonumber\\
&  =\gamma\left(  1+\mathbf{v}/c\right)  , \label{udef}%
\end{align}
where $\gamma=dt/d\tau$ is its \emph{time-dilation} factor, and $\mathbf{v}%
=d\mathbf{r}/dt$ is its coordinate velocity. Since $cd\tau$ is an invariant
interval in spacetime, $u$ transforms in the same way as $dr$ and is by
definition \emph{unimodular,} that is of unit length:%
\begin{equation}
u\bar{u}=1. \label{unimod}%
\end{equation}
It follows immediately that $u$ and $\bar{u}$ are inverses of each other and
that%
\begin{equation}
\gamma=\frac{dt}{d\tau}=\left[  1-\left(  \frac{\mathbf{v}}{c}\right)
^{2}\right]  ^{-1/2}.
\end{equation}

A simple Lorentz rotation is a rotation in a single spacetime plane of two
dimensions. Any Lorentz rotation can be built up of products of such simple
rotations. If the rotation plane comprises two spatial directions, the
transformation is a common spatial rotation. If, instead, one of the
directions in the rotation plane is the scalar time axis, the transformation
is a boost (velocity transformation). A simple rotation mixes the components
of the spacetime vector in the rotation plane and leaves components
perpendicular to the plane unchanged.

The simple rule for calculating the Lorentz rotation that boosts the frame of
a paravector $p=p^{0}+\mathbf{p}$ from rest to proper velocity $u$ is
\begin{equation}
p\rightarrow p^{\prime}=up^{\triangle}+p^{\bot} \label{transf}%
\end{equation}
where $p^{\triangle}\ $is the part of $p$ that lies coplanar with the rotation
plane in spacetime and $p^{\bot}$ is the remaining part, namely the part
orthogonal to the rotation plane:%
\begin{equation}
p^{\triangle}=p-p^{\bot}=p^{0}+\left(  \mathbf{p\cdot\hat{v}}\right)
\mathbf{~\hat{v}},
\end{equation}
where $\mathbf{\hat{v}}$ is the unit vector in the direction of $\mathbf{v.}$
The spacetime plane of the Lorentz rotation contains the time axis,
$e_{0}\equiv1$, the boost direction $\mathbf{v,}$ and all linear combinations
of $e_{0}$ and $\mathbf{v.}$ The part of $p$ in this plane is simply
multiplied by $u.$ The part $p^{\bot}$ perpendicular to the rotation plane,
with components on the spatial directions perpendicular to $\mathbf{v,}$ is
unchanged by the transformation.

The transformation (\ref{transf}) is all one needs to boost any spacetime
vector. As shown below, its form is the same as for spatial rotations. The
inverse transformation is given by replacing $u$ by $\bar{u},$ as seen from
(\ref{unimod}). No matrices or tensors are needed. To evaluate the boost
(\ref{transf}), only the products of scalars and the geometric product of
collinear vectors is needed, and as we saw above, the geometric product of
collinear vectors is just their scalar product.

In particular, this transformation applies to the basis vectors of paravector
space: Suppose our boost is along $e_{1},$ that is the proper velocity of the
boost is $u=\gamma\left(  1+ve_{1}/c\right)  .$ Then the transformation%
\begin{align*}
e_{0}  &  \equiv1\rightarrow u\\
e_{1}  &  \rightarrow ue_{1}\\
e_{2}  &  \rightarrow e_{2},\ e_{3}\rightarrow e_{3}%
\end{align*}
gives the spacetime basis of the frame moving with proper velocity $u.$ This
is easily plotted on a spacetime diagram if we note%
\begin{align*}
ue_{0}  &  =u=\gamma\left(  e_{0}+\frac{v}{c}e_{1}\right) \\
ue_{1}  &  =\gamma\left(  e_{1}+\frac{v}{c}e_{0}\right)  .
\end{align*}
%

\begin{figure}
[tbh]
\begin{center}
\includegraphics[
height=1.8524in,
width=2.2952in
]%
{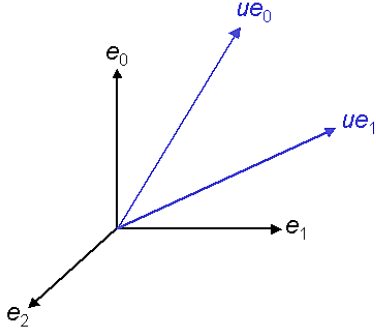}%
\caption{Boost of the basis for $\mathbf{v}=0.6c\mathbf{e}_{1}.$}%
\label{stdiag}%
\end{center}
\end{figure}
For example, if $v=0.6c,$ then $\gamma=5/4$ and the basis paravectors of the
moving frame are%
\begin{align*}
ue_{0} &  =\frac{5}{4}e_{0}+\frac{3}{4}e_{1}\\
ue_{1} &  =\frac{3}{4}e_{0}+\frac{5}{4}e_{1}%
\end{align*}
(see Fig. \ref{stdiag}). The perpendicular vectors $e_{2}$ and $e_{3}$ (not
shown) are unchanged.

\subsection{Spacetime Geometry}

\begin{figure}
[tbh]
\begin{center}
\includegraphics[
height=2.4448in,
width=2.6273in
]%
{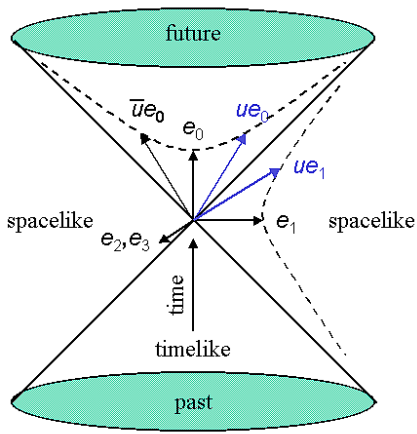}%
\caption{Geometry of spacetime}%
\label{spacetime}%
\end{center}
\end{figure}

It is worthwhile to point out some surprising features of spacetime geometry.
Note first that the transformed paravectors have the same square length as the
original ones:%
\begin{align*}
ue_{0}\left(  \overline{ue_{0}}\right)   &  =u\bar{u}=1\\
ue_{1}\left(  \overline{ue_{1}}\right)   &  =ue_{1}\bar{e}_{1}\bar{u}%
=-u\bar{u}=-1~.
\end{align*}
Obviously the defined square length of spacetime vectors does not correspond
to the Euclidean length that would be measured with a ruler on a diagram such
as Fig. \ref{stdiag}. The Euclidean length of a spacetime vector $p$ is
$\left\langle p^{2}\right\rangle _{S}^{1/2}.$ A good exercise for students is
to work out the loci of spacetime vectors of square length $\pm1$ on a
spacetime diagram. The result gives hyperboloids of revolution as seen in Fig.
\ref{spacetime}, which are asymptotic to the lightcone $r\bar{r}=0$. Note that
one spatial dimension in the $e_{2}e_{3}$ plane has been suppressed in the
diagram, and the lightcone, which is drawn as a two-dimensional surface,
actually represents the three-dimensional \emph{hypersurface} of a light pulse
emitted at the origin.%

All space at an instant in time for an observer at rest is a hypersurface that
appears as a horizontal plane in the diagram. The intersection of such a
hypersurface with the lightcone gives a circle on the diagram that represents
a spherical surface in space. Since Lorentz rotations preserve the square
lengths of paravectors, they leave the paravectors within defined regions of
spacetime. Spacetime vectors can be classified as either past timelike, future
timelike, spacelike, or lightlike (null), and Lorentz rotations do not change
the classification.

Note further that the paravectors $ue_{0}$ and $ue_{1}$ are orthogonal:%
\[
\left\langle ue_{1}\overline{ue}_{0}\right\rangle _{S}=\left\langle \bar
{u}u\bar{e}_{1}\right\rangle _{S}=0,
\]
and a spacetime vector on the lightcone is orthogonal to itself. Orthogonal
paravectors are not generally perpendicular (in a Euclidean sense) on the
spacetime diagram, but $ue_{1}$ is perpendicular to $\overline{ue_{0}}.$ More
generally, paravectors $p,q$ are perpendicular (on the diagram) if
$\left\langle pq\right\rangle _{S}=0,$ and orthogonal paravectors are always
perpendicular to each other's Clifford conjugate.

\subsection{Examples: Lorentz Contraction and Velocity Composition}

The time dilation derived above implies Lorentz contraction of a moving
object. As an example, consider a car racing at constant velocity
$\mathbf{v}=v\mathbf{e}_{1}$ in the lab. We place a clock at the origin and
measure the time for the car to cross the line $\mathbf{x\cdot e}_{1}=0$. If
$l$ is the length of the car in the lab, it takes a time $t=l/v$ for the car
to cross. In the frame of the car, the clock has velocity $-\mathbf{v}$ and
the crossing time is therefore dilated to $t_{0}=\gamma t,$ implying a longer
car length $l_{0}=vt_{0}=\gamma l.$ The car length in the lab is said to be
\emph{Lorentz contracted} by the factor $\gamma^{-1}$ relative to that in the
frame of the car. It is important to specify the motion of the clock that
records the events. It is easy for students to forget that the synchronization
of clocks at different positions is generally destroyed by a boost. Because
boosts are so easily calculated in APS, students can readily work out the full details:

In the lab, the spacetime position of the front of the moving car is
$x_{1}=ct+\mathbf{v}t$ and the rear of the car is at $x_{2}=x_{1}%
-l\mathbf{e}_{1},$ where $l$ is the length of the car in the lab. The rear of
the car crosses the start line when $\mathbf{x}_{2}\cdot\mathbf{e}_{1}=0,$
that is when $t=l/v.$ In the frame of the car, the spacetime positions are
found by boosting the car to rest, using the proper velocity $\bar{u}:$%
\begin{align*}
\bar{u}x_{1}  &  =\bar{u}\left(  c+\mathbf{v}\right)  t=\gamma^{-1}ct\\
\bar{u}x_{2}  &  =\bar{u}\left(  x_{1}-l\mathbf{e}_{1}\right)  =\gamma
^{-1}ct+\gamma vl/c-\gamma l\mathbf{e}_{1}.
\end{align*}
The vector part of $\bar{u}\left(  x_{1}-x_{2}\right)  $ gives the length of
the car in its rest frame, namely $\gamma l,$ which is larger than that in the
lab by the factor $\gamma.$

While the scalar parts, and hence lab times, of $x_{1},x_{2}$ are equal, the
times in the frame of the car differ by $\gamma vl/c^{2}.$ This shows that
clocks synchronized in the lab are not synchronized in the rest frame. The
scalar part of $\bar{u}x_{2}$ gives the time at the rear of the car in its
rest frame, and by substituting $l=vt$ we get the time of the rear crossing%
\[
\gamma^{-1}t+\gamma vl/c^{2}=\gamma t\left(  \gamma^{-2}+v^{2}/c^{2}\right)
=\gamma t
\]
in the car frame, which is just the dilated time of the clock fixed in the lab.

As an even simpler application, consider the composition of two collinear
boosts $u_{BA}$ and $u_{CB}$ ($u_{BA}$ can be read as the proper velocity of
$A$ with respect to $B$, etc.). Since the proper velocity $u_{BA}$ is itself a
spacetime vector, we can use (\ref{transf}) with $u$ given by $u_{CB}$ to
obtain simply the product (note that $\mathbf{u}_{BA}^{\bot}=0$)%
\begin{equation}
u_{CA}=u_{CB}u_{BA}~. \label{uprod}%
\end{equation}
Thus the \emph{addition }of collinear velocities in Galilean transformations
becomes a \emph{product }of proper velocities in relativity. The vector part
of (\ref{uprod}) gives%
\begin{equation}
\gamma_{CA}\mathbf{v}_{CA}=\gamma_{CB}\gamma_{BA}\left(  \mathbf{v}%
_{CB}+\mathbf{v}_{BA}\right)  \label{uuV}%
\end{equation}
and the scalar part is%
\begin{equation}
\gamma_{CA}=\gamma_{CB}\gamma_{BA}\left(  1+\mathbf{v}_{CB}\cdot
\mathbf{v}_{BA}/c^{2}\right)  . \label{uuS}%
\end{equation}
Their ratio gives the standard result immediately:%
\begin{equation}
\mathbf{v}_{CA}=\frac{\mathbf{v}_{CB}+\mathbf{v}_{BA}}{1+\mathbf{v}_{CB}%
\cdot\mathbf{v}_{BA}/c^{2}},
\end{equation}
which demonstrates that at speeds small compared to $c,$ the Galilean result
$\mathbf{v}_{CA}=\mathbf{v}_{CB}+\mathbf{v}_{BA}$ is obtained.

The composition of non-collinear velocities, while not so commonly given, is
easily found from (\ref{transf}):%
\begin{align*}
u_{CA} &  =u_{CB}u_{BA}^{\triangle}+u_{BA}^{\bot}\\
&  =\gamma_{CB}\left(  1+\frac{\mathbf{v}_{CB}}{c}\right)  \gamma_{BA}\left(
1+\frac{\mathbf{v}_{BA}^{\Vert}}{c}\right)  +\gamma_{BA}\frac{\mathbf{v}%
_{BA}^{\bot}}{c},
\end{align*}
where $\mathbf{v}_{BA}^{\Vert}=\mathbf{v}_{BA}-\mathbf{v}_{BA}^{\bot}$ is the
component of $\mathbf{v}_{BA}$ along $\mathbf{v}_{CB}~.$ The scalar part is as
before (\ref{uuS}) but the vector part (\ref{uuV}) is modified to%
\[
\gamma_{CA}\mathbf{v}_{CA}=\gamma_{CB}\gamma_{BA}\left(  \mathbf{v}%
_{CB}+\mathbf{v}_{BA}^{\Vert}\right)  +\gamma_{BA}\mathbf{v}_{BA}^{\bot},
\]
giving%
\[
\mathbf{v}_{CA}=\frac{\mathbf{v}_{CB}+\mathbf{v}_{BA}^{\Vert}+\mathbf{v}%
_{BA}^{\bot}/\gamma_{CB}}{1+\mathbf{v}_{CB}\cdot\mathbf{v}_{BA}/c^{2}}%
\]
See also Subsection V.2 on extensions to rotors.

\subsection{Spatial Rotations and Bivectors}

To better understand why boosts are considered rotations in spacetime and why
they have the algebraic form (\ref{transf}), we compare them to common
rotations in physical space. Such spatial rotations are Lorentz rotations in a
spatial plane, and they have exactly the same form as (\ref{transf}) but with
$u$ replaced by a rotation element such as%
\begin{equation}
\cos\theta+\mathbf{e}_{2}\mathbf{e}_{1}\sin\theta=\exp\left(  \theta
\mathbf{e}_{2}\mathbf{e}_{1}\right)  ,\label{Euler}%
\end{equation}
which gives a rotation by the angle $\theta$ in the plane $\mathbf{e}%
_{2}\mathbf{e}_{1},$ the plane that contains all linear combinations of
$\mathbf{e}_{1}$ and $\mathbf{e}_{2}.$ The Euler-like relation (\ref{Euler})
follows by power-series expansion and the easily verified result that $\left(
\mathbf{e}_{2}\mathbf{e}_{1}\right)  ^{2}=-1.$ Any product of orthogonal
vectors is called a \emph{bivector} and is an intrinsic representation of the
plane spanned by the vector factors. Bivectors generate spatial rotations, and
with their help, such rotations can be evaluated with simple algebra instead
of matrices. Recall from (\ref{innerprod}) that perpendicular vectors
anticommute. The bivectors $\mathbf{e}_{1}\mathbf{e}_{2}$ and $\mathbf{e}%
_{2}\mathbf{e}_{1}$ thus differ by a sign, which may be thought of as
indicating the circulation pattern or rotation direction in the plane. Since
$\left(  \mathbf{e}_{1}\mathbf{e}_{2}\right)  \mathbf{e}_{1}=-\mathbf{e}_{2}$
and $\left(  \mathbf{e}_{1}\mathbf{e}_{2}\right)  \mathbf{e}_{2}%
=\mathbf{e}_{1},$ the product of $\mathbf{e}_{1}\mathbf{e}_{2}$ from the left
with any linear combination $\mathbf{v}=v_{x}\mathbf{e}_{1}+v_{y}%
\mathbf{e}_{2}$ rotates $\mathbf{v}$ clockwise in the $\mathbf{e}%
_{1}\mathbf{e}_{2}$ plane by $90^{\circ}.$ If $\mathbf{e}_{1}\mathbf{e}_{2}$
is replaced by its inverse, $\mathbf{e}_{2}\mathbf{e}_{1},$ the rotation is
counter-clockwise by $90^{\circ}.$

For example, to rotate the paravector
\begin{equation}
p=p^{0}e_{0}+p_{x}\mathbf{e}_{1}+p_{y}\mathbf{e}_{2}+p_{z}\mathbf{e}_{3}%
\end{equation}
by the angle $\theta$ in the $\mathbf{e}_{2}\mathbf{e}_{1}$ plane, we
calculate%
\begin{align}
p  &  \rightarrow\exp\left(  \mathbf{e}_{2}\mathbf{e}_{1}\theta\right)
p^{\triangle}+p^{\bot}\label{rotate}\\
&  =\left(  \cos\theta+\mathbf{e}_{2}\mathbf{e}_{1}\sin\theta\right)  \left(
p_{x}\mathbf{e}_{1}+p_{y}\mathbf{e}_{2}\right)  +p^{0}e_{0}+p_{z}%
\mathbf{e}_{3}.\nonumber
\end{align}
The form of the transformation (\ref{transf}) ensures that components
orthogonal to the plane of rotation are unchanged. Here, components along
$\mathbf{e}_{0}$ and $\mathbf{e}_{3}$ are invariant while those in the plane
are rotated:%
\begin{align*}
p_{x}\mathbf{e}_{1}+p_{y}\mathbf{e}_{2}  &  \rightarrow\left(  p_{x}%
\mathbf{e}_{1}+p_{y}\mathbf{e}_{2}\right)  \cos\theta+\left(  p_{x}%
\mathbf{e}_{2}-p_{y}\mathbf{e}_{1}\right)  \sin\theta\\
&  =\left(  p_{x}\cos\theta-p_{y}\sin\theta\right)  \mathbf{e}_{1}+\left(
p_{y}\cos\theta+p_{x}\sin\theta\right)  \mathbf{e}_{2}~.
\end{align*}

\section{Duals and Cross Products}

The bivector $\mathbf{e}_{1}\mathbf{e}_{2}$ plays a role similar to that of
the unit imaginary for rotations in the $\mathbf{e}_{1}\mathbf{e}_{2}$ plane.
There is a difference, however, in that $\mathbf{e}_{1}\mathbf{e}_{2}$ is seen
to anticommute with vectors in the plane:%
\[
\mathbf{e}_{1}\mathbf{e}_{2}p^{\triangle}=-p^{\triangle}\mathbf{e}%
_{1}\mathbf{e}_{2}~.
\]
It is the \emph{volume element} $\mathbf{e}_{1}\mathbf{e}_{2}\mathbf{e}_{3}$
in APS that can be identified to $i:$ it squares to $-1$ and commutes with all
vectors and their products. With the identification%
\[
\mathbf{e}_{1}\mathbf{e}_{2}\mathbf{e}_{3}=i,
\]
every bivector in APS is equivalent to an imaginary vector directed
perpendicular to the plane of the bivector. For example%
\[
\mathbf{e}_{1}\mathbf{e}_{2}=\mathbf{e}_{1}\mathbf{e}_{2}\mathbf{e}%
_{3}\mathbf{e}_{3}=i\mathbf{e}_{3}.
\]
The vector $\mathbf{e}_{3}$ is said to be \emph{dual} to the plane
$\mathbf{e}_{1}\mathbf{e}_{2}.$ It is common in physics to represent a spatial
plane by its dual vector. Indeed, this is the principal use of the vector
cross product since the vector dual to a plane is the cross product of vectors
in the plane. In the case above, $\mathbf{e}_{3}=\mathbf{e}_{1}\times
\mathbf{e}_{2}~.$ However, the dual vector is an \emph{extrinsic
}representation of the plane: it depends on the space in which the plane
resides. A two-dimensional plane in a space with four or more dimensions does
not possess a unique dual vector. The bivector, as mentioned above, is an
\emph{intrinsic} representation of a plane: it depends only on two linearly
independent vectors in the plane.

More generally, the bivector part of the product of vectors $\mathbf{p,q}$ is
related to the cross product by%
\begin{equation}
i\mathbf{p\times q=}\frac{1}{2}\left(  \mathbf{pq-qp}\right)  =\mathbf{pq}%
-\mathbf{p\cdot q} \label{bivector}%
\end{equation}
The axis of rotation is the vector dual to the rotation plane. Such a vector
is called an \emph{axial vector} (or pseudovector), denoting its invariance
under such rotations as well as under a spatial inversion of its vector
factors. Important examples are the orbital angular momentum
$\mathbf{L=r\times p}$ and the magnetic field of a moving charge, both of
which are vectors dual to the plane of motion (the plane of $\mathbf{r}$ and
$\mathbf{p}$). By using the dual vector, one can write the rotation element
$\exp\left(  \mathbf{e}_{2}\mathbf{e}_{1}\theta\right)  $ in terms of the
rotation axis: $\exp\left(  -i\mathbf{e}_{3}\theta\right)  .$

A geometrically distinct use of the cross product occurs in the cross product
of a vector with a pseudovector, for example $\left(  \mathbf{p\times
q}\right)  \times\mathbf{r,}$ which arises algebraically as the product of the
bivector (\ref{bivector}) with the coplanar component $\mathbf{r}^{\triangle}$
of $\mathbf{r.}$ As seen above, the result is a vector in the plane of
$\mathbf{p}$ and $\mathbf{q}$ that is perpendicular to $\mathbf{r}^{\triangle
}$ (and hence to $\mathbf{r}$). A common example of this use of the cross
product occurs in the Lorentz force $q\mathbf{v\times B~.}$ These two
geometrical uses of the cross product can be the source of confusion in usual
vector treatments, but they are cleanly distinguished in APS, where vectors
and bivectors are distinct elements.

Aside: One can identify $\frac{1}{2}\left(  \mathbf{pq-qp}\right)  $ as the
exterior or wedge product $\mathbf{p\wedge q}$ of the vectors $\mathbf{p,q}.$
This product is important in treatments with differential forms and in
geometric algebras of higher dimension. However, its use requires further
rules for combining wedge and dot products with other elements, a complication
that can be avoided in APS. (See also Section VI.)

\subsection{Boosts as Spacetime Rotations}

The relation of spatial rotations to boosts (\ref{transf}) is strengthened by
writing the proper velocity of the boost in the explicitly unimodular form%
\begin{equation}
u=\exp\left(  w\mathbf{\hat{v}}\right)  =\cosh w+\mathbf{\hat{v}}\sinh
w,\label{expw}%
\end{equation}
where the boost parameter $w$ is called the \emph{rapidity }of the boost. A
comparison with the defining form (\ref{udef}) establishes that $\gamma=\cosh
w.$ The unit vector $\mathbf{\hat{v}}=\mathbf{\hat{v}}\bar{e}_{0}$ represents
the spacetime plane of rotation for the boost, namely the plane containing all
real linear combinations of the direction $\mathbf{\hat{v}}$ of the boost
velocity and the time axis $e_{0}.$ It is the product of orthogonal directions
in spacetime. We discuss spacetime planes more thoroughly below in Section V.
The essential difference between boosts and spatial rotations arises from
geometrical differences in the planes of rotation. The rotation plane for
boosts includes the time axis $e_{0}$ and its generator (such as
$\mathbf{\hat{v}}\bar{e}_{0}$) squares to $+1.$ Spatial rotations, on the
other hand, are generated by a bivector that squares to $-1.$ Because of this
difference, the parameter $\theta$ for the spatial rotation is periodic, with
a rotation by angle $\theta+2n\pi$ for any integer $n$ giving the same result
as one by $\theta.$ All possible rotations in the plane are given by a finite
range of the parameter, say $0\leq\theta<2\pi.$ However, for boosts, each
value of the parameter in the range $-\infty<w<\infty$ gives a distinct boost.
Boosts cannot rotate paravectors through the light cone; they can only tilt
them within their spacelike or timelike regions (see Fig. \ref{spacetime}).

\subsection{Electromagnetic Field}

\begin{figure}
[tbh]
\begin{center}
\includegraphics[
height=2.0972in,
width=2.3419in
]%
{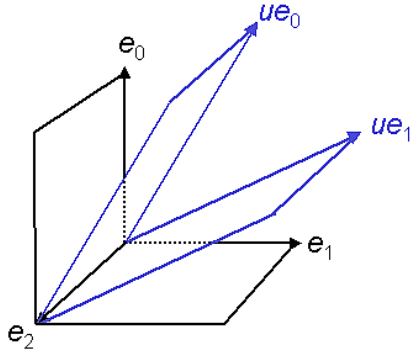}%
\caption{Boosts of spacetime planes are determined by boosts of the spacetime
vectors in them.}%
\label{stplanes}%
\end{center}
\end{figure}

An introductory physics class may also need to transform electromagnetic
fields. Under spatial rotations, electric and magnetic fields are transformed
exactly like other vectors. However, boosts act differently on electromagnetic
fields than on spacetime vectors because the fields transform as spacetime
planes rather than spacetime vectors. As above, by \textquotedblleft
planes\textquotedblright\ we mean two-dimensional geometrical objects that
contain all real linear combinations of two noncollinear paravectors. The
boost transformation for spacetime planes can be found from that for
paravectors, as illustrated in Fig. \ref{stplanes}.%

The electric and magnetic fields are given by the real and imaginary parts of
a covariant electromagnetic field $\mathbf{F=E}+ic\mathbf{B}$, a complex
vector that represents a spacetime plane at a point in spacetime. The real
part of $\mathbf{F}$ is the electric field, and its imaginary part is the
bivector for a spatial plane whose normal vector is the magnetic field
$\mathbf{B}$ (times $c$ in SI units). A static electric field along
$\mathbf{e}_{2}$ sweeps out a spacetime plane containing $\mathbf{e}_{2}$ and
$e_{0}.$ Under a boost along $\mathbf{e}_{1},$ this spacetime plane is n in
the plane of $\mathbf{e}_{1}$ and $e_{0},$ thereby picking up a spatial
(horizontal) component in the $\mathbf{e}_{1}\mathbf{e}_{2}$ plane that
represents a magnetic field in the $\mathbf{e}_{3}$ direction. Similarly, a
magnetic field along $\mathbf{e}_{3},$ corresponding to a component of the
electromagnetic field in the $\mathbf{e}_{1}\mathbf{e}_{2}$ plane, is rotated
by the boost to give a vertical-plane component corresponding to an electric
field along $\mathbf{e}_{2}.$ More generally, under a boost to proper velocity
$u,$ $\mathbf{F}$ can be shown to transform as%
\begin{equation}
\mathbf{F}\rightarrow u\mathbf{F}^{\bot}+\mathbf{F}^{\Vert}\label{Fboost}%
\end{equation}
where $\mathbf{F}^{\Vert}=\mathbf{F\cdot\hat{v}~\hat{v}}=\mathbf{F-F}^{\bot}$
is the component of $\mathbf{F}$ along the boost direction. Note that the term
$u\mathbf{F}^{\bot}$ is the sum of $\gamma\mathbf{vE}^{\bot},$ a bivector that
contributes to the magnetic field, and $i\gamma\mathbf{vB}^{\bot}%
=-i\gamma\mathbf{Bv}^{\triangle},$ a vector that contributes to the electric
field. Here $\mathbf{v}^{\triangle}$ is the component of $\mathbf{v}$ in the
plane of $i\mathbf{B}$ and hence perpendicular to $\mathbf{B,}$ and
$-i\gamma\mathbf{Bv}^{\triangle}$ lies in the plane of $i\mathbf{B}$ and is
perpendicular to $\mathbf{v.}$ The full significance of the spacetime form of
the field and a derivation of its transformation can be left for later (see
Section V).

This completes a major goal of this paper: to show how arbitrary boosts and
spatial rotations can be easily calculated algebraically, without matrices or
tensors. Instructors can readily add the many other examples traditionally
presented in first courses on relativity, all without matrices or tensors. The
formalism maintains close contact with vectors and with the geometry of
physical space implicit in the vector notation, and by simply adding time
components as scalars to the vectors, it provides an intuitive introduction to
relativity suitable for use early in the physics curriculum.

\section{Extensions}

APS does considerably more than allow relativistic computations with vectors:
it efficiently models the \emph{geometry of spacetime}, representing objects
in relativity covariantly as paravectors and their products. We saw above how
paravectors in APS unite the space and time components of spacetime vectors.
Boosts tilt the paravectors and thereby mix space and time parts.

The material that follows goes beyond what would normally be presented to
beginning students, but bright students and instructors may want a fuller
understanding of the algebraic formalism as a framework for treating advanced
problems in physics. First the APS treatment of planes is extended to
spacetime, and then Lorentz transformations are generalized to rotations in
arbitrary spacetime planes. Next spin transformations are introduced, and a
new decomposition for compound rotations is given. Finally transformations are
derived for objects other than spacetime vectors, in particular, for
electromagnetic fields, and applied to plane waves to reveal a curious feature
of spacetime geometry.

\subsection{Planes in Spacetime}

The important geometrical concept of two-dimensional planes in which rotations
occur is readily extended to the four dimensions of spacetime. Whereas spatial
planes are represented by bivectors, that is, a product of perpendicular
vectors, planes in spacetime are represented by \emph{biparavectors,} that is
products of two orthogonal paravectors. The plane spanned by orthogonal
paravectors $p$ and $q$ is represented by $p\bar{q}$ or $q\bar{p}.$ These
biparavectors are written as operators that rotate paravectors in the plane
into an orthogonal direction in the plane, and they differ by the sense of the
rotation. Thus, $p\bar{q},$ when multiplying $q$ from the left, rotates $q$
into the direction $\pm p$ and $p$ into the direction $\pm q$ whereas
$q\bar{p}=-p\bar{q}$ rotates in the opposite direction. (The actual signs
depend on the signs of $\bar{q}q$ and $\bar{p}p,$ respectively.)

There are $\binom{4}{2}=6$ linearly independent planes in spacetime, and a
suitable orthonormal basis of biparavectors is $\left\{  e_{1}\bar{e}%
_{0},e_{2}\bar{e}_{0},e_{3}\bar{e}_{0},e_{3}\bar{e}_{2},e_{1}\bar{e}_{3}%
,e_{2}\bar{e}_{1}\right\}  .$ Each basis biparavector generates rotations in a
spacetime plane. The first three generate boosts, the last three generate
spatial rotations. By using $e_{1}e_{2}e_{3}=i=e_{0}\bar{e}_{1}e_{2}\bar
{e}_{3},$ we can also express the biparavector basis as $\left\{  e_{1}%
,e_{2},e_{3},ie_{1},ie_{2},ie_{3}\right\}  ,$ which comprises three real and
three imaginary vectors in dual pairs. We saw above that imaginary vectors
represent spatial planes in terms of their dual (or axial) vectors. The real
vectors can be thought of here as \emph{persistent vectors,} ones that persist
in time and sweep out timelike planes (planes that contain the time axis
$e_{0}=1$) in spacetime. In spacetime, the timelike plane $e_{j}=e_{j}\bar
{e}_{0}$ is dual to the spatial plane $ie_{j}.$ A plane in spacetime can
therefore have up to six independent components and be represented by a
complex vector.

An important physical plane in spacetime is the electromagnetic field
$\mathbf{F}$ at some spacetime point $r.$ If $\mathbf{F}$ is expanded in the
biparavector basis, the coefficients give the usual tensor components
$F^{\mu\nu}:$%
\begin{equation}
\mathbf{F}=\mathbf{E}+ic\mathbf{B}=\frac{1}{2}\sum_{\mu,\nu=0}^{3}F^{\mu\nu
}\left\langle e_{\mu}\bar{e}_{\nu}\right\rangle _{V},\label{Ffield}%
\end{equation}
where $\left\langle e_{\mu}\bar{e}_{\nu}\right\rangle _{V}=\frac{1}{2}\left(
e_{\mu}\bar{e}_{\nu}-e_{\nu}\bar{e}_{\mu}\right)  $ is the vector-like part of
$e_{\mu}\bar{e}_{\nu}$; it vanishes when $\mu=\nu$ but otherwise equals
$e_{\mu}\bar{e}_{\nu}.$ The expansion (\ref{Ffield}) relates the tensor
components $F^{\mu\nu}$ of the electromagnetic field $\mathbf{F}$ to local
vector components of $\mathbf{E}$ and $\mathbf{B.}$ The factor of $\frac{1}%
{2}$ compensates for the appearance of each basis biparavector twice. We have
already noted that $\mathbf{B}$ is the vector dual to spatial planes. The
vector $\mathbf{E}$ is an example of a vector that persists in time; it sweeps
out the timelike plane $\mathbf{E=E}\bar{e}_{0}$ in time. For any observer,
$\mathbf{F}$ may have both a timelike projection $\mathbf{E}$ and a spatial
one $ic\mathbf{B.}$ We should note that although we have spoken of
$\mathbf{F}$ as a single spacetime plane, in four dimensions it is also
possible for $\mathbf{F}$ to be the sum of two orthogonal planes, in which
every paravector in one is orthogonal to every paravector in the other. In
contrast to spatial planes in three dimensions, where two planes always share
a common vector and can be added to give a single spatial plane, orthogonal
planes in four dimensions are dual to each other and cannot be combined into a
single real plane. However, since dual spacetime objects in APS are related by
the volume element $e_{0}\bar{e}_{1}e_{2}\bar{e}_{3}=i,$ any sum of spacetime
planes in APS can be expressed as a complex scalar times a single plane.

\subsection{Rotors and the Group of Lorentz Rotations}

The transformations (\ref{transf}) and (\ref{rotate}) allow simple
calculations of individual boosts and spatial rotations, but they are not
particularly convenient for representing a sequence or group of such
transformations. Instead, the general Lorentz rotation of paravector $p$ can
be expressed as a \emph{spin transformation}\cite{Loun97}%
\begin{equation}
p\rightarrow LpL^{\dag},\label{Lorrot}%
\end{equation}
where the \emph{Lorentz rotor} can be written $L=\pm\exp\left(  \mathbf{W}%
/2\right)  .$ For a \emph{simple }Lorentz rotation, that is, a rotation in a
single spacetime plane, $L=+\exp\left(  \mathbf{W}/2\right)  ,$ and
$\mathbf{W}$ is a biparavector that gives both the plane and the magnitude of
the rotation. Compound rotations, that is simultaneous rotations in orthogonal
planes, can be expressed as a commuting product of two simple rotations. The
dagger $\dag$ indicates reversion, that is the reversal of the order of all
vector factors; it is equivalent to hermitian conjugation for any matrix
representation in which the basis vectors are hermitian. It is easily seen
that $L^{\dag}=\exp\left(  \mathbf{W}^{\dag}/2\right)  $ and that for spatial
rotations, $\mathbf{W}$ is \textquotedblleft imaginary\textquotedblright, that
is $\mathbf{W}^{\dag}=-\mathbf{W,}$ whereas for boosts $\mathbf{W}$ is
\textquotedblleft real\textquotedblright: $\mathbf{W}^{\dag}=\mathbf{W.}$
Since $\mathbf{\bar{W}}=-\mathbf{W,}$ $L$ is unimodular: $L\bar{L}=1.$

We can easily show that the spin transformation (\ref{Lorrot}) reduces to the
form (\ref{transf}) and (\ref{rotate}) for any simple Lorentz rotation. Let
$\mathbf{W}$ be the product $r\bar{s}$ of real orthogonal paravectors. Then,
for any paravector $\alpha r+\beta s$ in the plane of $\mathbf{W,}$ where
$\alpha,\beta$ are real scalars,%
\begin{equation}
\mathbf{W}\left(  \alpha r+\beta s\right)  =\left(  \alpha r+\beta s\right)
\mathbf{W}^{\dag},
\end{equation}
with $\mathbf{W}^{\dag}=\left(  r\bar{s}\right)  ^{\dag}=\bar{s}^{\dag}%
r^{\dag}=\bar{s}r.$ It follows from multiple applications of this relation in
the power-series expansion of $L$ that%
\begin{equation}
L\left(  \alpha r+\beta s\right)  L^{\dag}=L^{2}\left(  \alpha r+\beta
s\right)  .
\end{equation}

On the other hand, if $q$ is any paravector orthogonal to both $r$ and $s$
(and hence to the plane containing $r$ and $s$), then since $q\bar{s}%
=-s\bar{q}$ and $\bar{q}r=-\bar{r}q,$%
\begin{equation}
q\mathbf{W}^{\dag}=q\bar{s}r=-s\bar{q}r=s\bar{r}q=-\mathbf{W}q~.
\end{equation}
It follows that orthogonal paravectors are invariant under rotations in the
plane:%
\begin{equation}
LqL^{\dag}=L\bar{L}q=q.
\end{equation}
Consequently, the simple Lorentz rotation (\ref{Lorrot}) of an arbitrary
paravector $p$ can always be split into two parts that transform distinctly:%
\begin{equation}
p\rightarrow L\left(  p^{\triangle}+p^{\bot}\right)  L^{\dag}=L^{2}%
p^{\triangle}+p^{\bot}.\label{Lrotate}%
\end{equation}
Our previous expressions (\ref{transf}) and (\ref{rotate}) of boosts and
spatial rotations are seen as special cases of (\ref{Lrotate}). From
(\ref{Lrotate}) we can establish an explicit relation for the Lorentz rotor in
terms of any non-null paravector $p$ in the plane of rotation and the
transformed result $r=LpL^{\dag}=L^{2}p$. Thus, as may be verified by
squaring,%
\begin{equation}
L=\left(  rp^{-1}\right)  ^{1/2}=\frac{\left(  p+r\right)  p^{-1}}%
{\sqrt{2\left\langle \left(  p+r\right)  p^{-1}\right\rangle _{S}}}.
\end{equation}
Note that $Lp$ lies along $p+r,$ which bisects $p$ and $r,$ and that the
square-root factor is required for the normalization $L\bar{L}=1.$

We can readily extend the result (\ref{Lrotate}) for paravector rotations in
single planes to compound rotations, in which rotations are made in a pair of
dual planes. Because the biparavectors for the dual planes commute, the
Lorentz rotor for any compound rotation can be factored into a pair of simple
rotations, one for each of the dual planes: $L=L_{1}L_{2}=L_{2}L_{1}.$
Similarly, any paravector can be uniquely split into components coplanar with
the two planes: $p=p^{\triangle_{1}}+p^{\triangle_{2}},$ where $p^{\triangle
_{1}}$ is coplanar with the plane of rotation of $L_{1}$ and orthogonal to its
dual, that is the plane of rotation of $L_{2},$ and \emph{vice versa }for
$p^{\triangle_{2}}$. The compound rotation of $p$ is then easily split as
follows:%
\begin{equation}
LpL^{\dag}=L_{1}L_{2}\left(  p^{\triangle_{1}}+p^{\triangle_{2}}\right)
L_{2}^{\dag}L_{1}^{\dag}=L_{1}^{2}p^{\triangle_{1}}+L_{2}^{2}p^{\triangle_{2}%
}. \label{compound}%
\end{equation}
The result (\ref{Lrotate}) for simple Lorentz rotations is the special case of
the compound case (\ref{compound}) when the exponent of one of the pair of
rotors vanishes.

The forms (\ref{Lorrot}) and (\ref{Lrotate}) of the Lorentz rotations are
equivalent, and while (\ref{Lrotate}) is often simpler to evaluate, the spin
form (\ref{Lorrot}) leads to powerful spinor methods with the Lorentz group.
An important example is the use of the spin form (\ref{Lorrot}) to extend
Lorentz rotations to products of paravectors. In particular, any product
$p\bar{q}$ of paravectors is seen to transform according to%
\begin{equation}
p\bar{q}\rightarrow\left(  LpL^{\dag}\right)  \left(  \bar{L}^{\dag}\bar
{q}\bar{L}\right)  =Lp\bar{q}\bar{L}~.
\end{equation}
The scalar part of $p\bar{q}$ is thus invariant under Lorentz transformations,
and any biparavector such as the electromagnetic field $\mathbf{F}$ transforms
as%
\begin{equation}
\mathbf{F}\rightarrow L\mathbf{F}\bar{L}~.
\end{equation}
Using methods analogous to those above, for simple rotors $L=\exp\left(
\mathbf{W}/2\right)  $ in non-null planes $\mathbf{W,}$ one can put this into
a form analogous to the paravector rotation (\ref{Lrotate}):%
\begin{equation}
\mathbf{F}\rightarrow L^{2}\left(  \mathbf{F-F}^{\mathbf{W}}\right)
+\mathbf{F}^{\mathbf{W}}\label{LrotField}%
\end{equation}
where $\mathbf{F}^{\mathbf{W}}=\left\langle \mathbf{FW}^{-1}\right\rangle
_{S}\mathbf{W}$ is the projection of $\mathbf{F}$ onto $\mathbf{W}$ and its
dual. This reduces to the forms (\ref{rotate}) and (\ref{Fboost}) for
rotations and boosts, respectively.

\subsection{Example: Boosts and Scaled Rotations}

Propagating plane waves are null fields ($\mathbf{F}^{2}=0$), which have the
form
\begin{equation}
\mathbf{F}=\left(  1+\mathbf{\hat{k}}\right)  \mathbf{E}=\mathbf{E}\left(
1-\mathbf{\hat{k}}\right)  \mathbf{,}%
\end{equation}
where the electric field $\mathbf{E}$ is perpendicular to the direction
$\mathbf{\hat{k}}$ of the energy flow. This form follows easily from Maxwell's
equations for source-free space applied to a field $\mathbf{F}$ that is
assumed to depend on spacetime location $x$ only through the scalar
$s=\left\langle k\bar{x}\right\rangle _{S}$, where $k$ is a constant
paravector.\cite{Bay99} It is found that $k$ is null and can thus be written%
\begin{equation}
k=\frac{\omega}{c}\left(  1+\mathbf{\hat{k}}\right)  ,
\end{equation}
where $c$ is the wave speed, $\mathbf{\hat{k}}$ is the propagation direction.
If the wave is monochromatic, $\omega$ is its frequency and the corresponding
photon momentum in spacetime is $\hbar k.$ We show here that any boost of such
a wave is equivalent to a rotation and dilation (rescaling), and that $k$ and
$\mathbf{F}$ are rotated and dilated by the same amount.

A boost of $\mathbf{F}$ and $k$ gives%
\begin{subequations}
\begin{align}
\mathbf{F}^{\prime} &  =L\mathbf{F}\bar{L}\\
k^{\prime} &  =LkL^{\dag}%
\end{align}
with a rotor%
\end{subequations}
\[
L=e^{w\mathbf{\hat{v}}/2}=\cosh\frac{w}{2}+\mathbf{\hat{v}}\sinh\frac{w}%
{2}=L^{\dag},
\]
where $w$ is the rapidity of the boost and $\mathbf{\hat{v}}$ its direction.
Now a curious property of null paravectors such as $1+\mathbf{\hat{k}}$ is
that they can \textquotedblleft gobble\textquotedblright\ (or
\textquotedblleft ungobble\textquotedblright) factors of $\mathbf{\hat{k}}$ in
what has become known as the \emph{pacwoman property:}\cite{Bay99}%
\begin{equation}
\mathbf{\hat{k}}\left(  1+\mathbf{\hat{k}}\right)  =1+\mathbf{\hat{k}=}\left(
1+\mathbf{\hat{k}}\right)  \mathbf{\hat{k}~.}%
\end{equation}
Consequently%
\begin{equation}
L\left(  1+\mathbf{\hat{k}}\right)  =\left(  \cosh\frac{w}{2}+\mathbf{\hat
{v}\hat{k}}\sinh\frac{w}{2}\right)  \left(  1+\mathbf{\hat{k}}\right)  ,
\end{equation}
and the factor $\cosh\frac{w}{2}+\mathbf{\hat{v}\hat{k}}\sinh\frac{w}{2}$ is a
scalar plus a bivector, which can always be factored into $e^{\alpha/2}R,$ the
product of a dilation factor $e^{\alpha/2}$ with a unitary rotor $R$ for a
spatial rotation in the plane of $\mathbf{\hat{v}}$ and $\mathbf{\hat{k}}$. In
the special case that $\mathbf{\hat{v}}\cdot\mathbf{\hat{k}}=0,$
$\mathbf{\hat{v}\hat{k}}$ is a spatial bivector and we have%
\begin{align}
L\left(  1+\mathbf{\hat{k}}\right)   &  =\frac{\cosh\frac{w}{2}}{\cos
\frac{\theta}{2}}\left(  \cos\frac{\theta}{2}+\mathbf{\hat{v}\hat{k}}\sin
\frac{\theta}{2}\right)  \left(  1+\mathbf{\hat{k}}\right)  \nonumber\\
&  =e^{\alpha/2}R\left(  1+\mathbf{\hat{k}}\right)  ,
\end{align}
where the rotor is%
\begin{equation}
R=e^{\theta\mathbf{\hat{v}\hat{k}}/2},
\end{equation}
the dilation factor is%
\begin{equation}
e^{\alpha/2}=\frac{\cosh\frac{w}{2}}{\cos\frac{\theta}{2}},
\end{equation}
and the rotation angle is given by%
\begin{equation}
\tan\frac{\theta}{2}=\tanh\frac{w}{2}%
\end{equation}
and thus by $\cos\theta=1/\cosh w=\gamma^{-1}=e^{-\alpha}.$ Combining these
results with their reversion and Clifford conjugation, we find that the boost
rotates and dilates both $\mathbf{F}$ and $k$ by the same amount:%
\begin{subequations}
\begin{align}
\mathbf{F}^{\prime} &  =e^{\alpha}R\mathbf{F}\bar{R}\\
k^{\prime} &  =e^{\alpha}RkR^{\dag}.
\end{align}

\section{Comparison to STA}

The formulation of relativistic physics used above is based on the algebra of
physical space (APS). The spacetime algebra (STA) is the geometric algebra
based on Minkowski spacetime. It was developed by Hestenes\cite{Hes66}, who
together with Doran and Lasenby\cite{DorLas03} and others have applied it to a
wide variety of problems in physics. In this section, we compare APS and STA
for introductory treatments of relativity.

APS and STA are closely related. Both are geometric algebras that emphasize
the geometric significance of vector products and avoid matrices and tensor
elements. The starting or ground space in APS is the space of vectors in
three-dimensional physical space, whereas STA is based on four-dimensional
vectors in Minkowski spacetime. In APS, spacetime vectors are represented by
real paravectors, inhomogeneous elements that are sums of scalars and vectors,
whereas in STA spacetime vectors are the homogeneous vectors of the ground
space. APS starts with vectors in a Euclidean metric and finds the Minkowski
metric as the natural metric of paravector space. In STA, the Minkowski metric
is assumed from the outset. In APS all the basis vectors can be taken as real
or hermitian. This is not possible in STA, where the metric requires either
the timelike basis vector to be real and the spacelike ones to be imaginary or
vice versa, depending on the metric signature. It is usual to assume a
negative metric signature in STA; the adoption of a Minkowski spacetime metric
with positive signature would create a nonequivalent algebra. In APS, the
signature of the paravector metric is trivially changed by reversing the sign
of the quadratic form for paravectors, that is, what one identifies as their
square lengths.

The number of real linearly independent elements in APS is 8, whereas in STA
it is 16. Any element in APS can be expressed as a complex paravector, where
the four parts, namely the real and imaginary scalar parts and the real and
imaginary vector parts represent distinct geometrical entities. APS and STA
are related by an isomorphism between APS and the even subalgebra of STA. As a
result, anything calculated in APS can also be calculated in STA. Less
obvious, it has recently been shown\cite{BaySob04} that any measurable
physical process that can be represented in STA can also be treated in APS.
Both APS and STA give covariant descriptions of relativistic phenomena, but in
STA the extra size also admits an absolute-frame model, in which each inertial
observer and each field or object frame is absolute, whereas in APS, only the
experimentally verifiable relative nature of spacetime vectors and frames is
posited. The relation of APS to physical-space vectors is immediate, and
different observers in APS see different splits of spacetime vectors into
scalar and vector parts as well as different splits of the electromagnetic
field into electric and magnetic parts. In STA, spacetime vectors and fields
can be absolute, and the measurable components arise by multiplying the
absolute object frame by the absolute observer time axis and extracting a
space/time split of the resulting bivector.

For introductory treatments, it is especially convenient that the volume
element in APS is the unit imaginary scalar, and that it is the same for both
physical space and for spacetime. The volume element $I$ in STA, on the other
hand, although it is called a pseudoscalar and is used to define dual
elements, it actually anticommutes with all vectors and thus acts rather like
a fifth dimension. The wedge product in APS is largely avoided, and indeed its
definition is problematic for inhomogeneous elements such as paravectors.
However, it is used extensively in STA, where one must pay attention to rules
for combining contractions (dot products) and wedge products and to the fact
that the introduction of $I$ into a product can change one type of product
into the other. An extra conjugation, the Clifford conjugation, is required in
APS but not STA. However, as seen above in Subsection III-C, the Clifford
conjugation provides a simple way to relate lengths and orthogonality in
Minkowski spacetime with measured (Euclidean) lengths and perpendicularity on
a spacetime diagram.

Some of the efficiency of APS in representing relativistic phenomena arises
from the double role playing of its element types. Thus a real scalar may be a
Lorentz invariant or it may be the time component of a spacetime vector, and a
real vector may be the spatial part of a spacetime vector or the real part of
a spacetime plane. Such ambiguity of roles may appear a potential source of
confusion, but in fact it is part of our language and common usage. The
context usually makes the role clear. For example, a proper-time interval
(times $c$) is both a Lorentz-invariant spacetime displacement and the time
component of the displacement in the particle rest frame. Similarly, the mass
of a particle (times $c$) is both the Lorentz invariant length of its
spacetime momentum and the time component of the momentum in the particle rest
frame. In APS, the same element plays both roles, but in STA, one
distinguishes, say, between the Lorentz invariant $mc$ and the time component
of the spacetime momentum, $mc\gamma_{0}~.$ Similarly, in APS the electric
field $\mathbf{E}$ is both a vector and, through its persistence in time, a
timelike plane $\mathbf{E}\bar{e}_{0}$ in spacetime, whereas in STA,
$\mathbf{E}$ and $\mathbf{E}\gamma_{0}$ remain distinct.

APS, in contrast to STA, provides a more direct extension of physical space
with fewer new rules and operators. The volume element, through which duality
is defined, is the same for three-dimensional physical space as for
four-dimensional paravector space, which represents spacetime. It is the true
pseudoscalar for both spaces in that it commutes with all elements of the
algebra. These distinctions argue for APS as the preferred vehicle for
teaching relativity in introductory physics. STA and, for that matter, the
differential-forms approach require more overhead and may be more easily
approached after students have been exposed to many of the concepts of
geometric algebra, and after they have mastered some of the spacetime geometry
from APS.

\section{Conclusion}

Hestenes\cite{Hes03a} has advocated geometric algebra as a universal language
of physics. The proposal made in the present article is a modest step in that
direction. The algebra of physical space, equivalent to the even half of
Hestenes' spacetime algebra,\cite{Hes03b} offers a simple approach to special
relativity that is suitable for introduction at an early stage into the
physics curriculum. As its name suggests, APS is based on physical
(three-dimensional Euclidean) space that is familiar to beginning students.
However, it also includes a four-dimensional linear space of paravectors,
which are formed by adding scalar time components to the vectors. The algebra
builds on the natural appearance of the Minkowski spacetime metric in
paravector space. Paravectors represent spacetime vectors, and their products
represent planes and other geometrical structures in spacetime. Rotations of
paravectors and their products give the physical Lorentz rotations. Only basic
elements of the algebra are needed for the student to easily calculate
arbitrary boosts and rotations without recourse to matrices or tensors. Space
and time are united in the spacetime continuum of paravector space, but
familiar spatial vectors and their physical significance are never left behind.

Other geometric algebras can also be used to formulate relativistic physics,
and STA has been especially well developed for this purpose. However, APS is
based more immediately on the familiar vectors of physical space, and it
requires only half as many independent elements as STA and avoids some of the
abstractions and potential hazards such as wedge products and noncommuting
pseudoscalars. Even Hestenes seems to agree that APS is the best algebra for
introducing relativity, since this is what he has used in the second edition
of his mechanics text.\cite{Hes99} A simpler alternative candidate is the
algebra of complex quaternions. It has a long
history\cite{Conway1912,Silberstein1912,Silberstein1914} as a mathematical
framework for relativity, but its broader use in the physics community has
remained limited, probably because the geometric interpretation is rather
disguised and it does not lend itself easily to a covariant formulation.

The APS approach is part of a comprehensive geometric algebra that nicely
displays both spatial and spacetime symmetries and can be applied to all areas
of relativistic physics. Much of the power of Clifford's geometric algebras
can be traced to their intrinsic representation of planes as bivectors, which
can generate rotations. Planes in the paravector space of APS are represented
by biparavectors. They not only generate Lorentz transformations but also
covariantly represent physical entities such as the electromagnetic field.
Some results of APS, such as analytic solutions for the relativistic motion of
charges in fields of propagating plane-wave pulses, are easy to obtain in the
algebra\cite{Bay99a} but almost impossible to find without use of its spinor
and projector tools. In this paper, I have illustrated the algebraic language
with several simple examples, a couple of new insights into understanding and
computing the geometry of spacetime in APS, and by proving that any boost of
the electromagnetic field of a propagating plane wave is equivalent to a
rotation and a dilation, and that the same rotation and dilation result for
both the field and its propagation (and thus momentum) paravector.

The successful introduction of relativity as an integral part of the
first-year physics curriculum faces difficulties regardless of the approach
taken. The concept of an observer-dependent spacetime in place of absolute
space and independent universal time may have a certain intuitive appeal, but
since that spacetime has an extra dimension and is no longer Euclidean, it
will challenge and stretch the minds of even the best students. Nevertheless,
that is the physical world as we have known it for almost a century, and it is
high time that we taught it directly to our beginning physics students. This
will only be practical if a suitable mathematical approach is used, one based
on the geometry of space and spacetime that avoids the unnecessary baggage of
tensor and matrix components. Proposing such an approach has been the primary
purpose of this paper. The challenge of relativistic concepts may be too great
for quantitative calculations in service courses to biologists or engineers,
say, but it may also be just the ingredient that excites good students to
major in physics. Certainly being able to recognize and use relativistic
symmetries will facilitate studies in areas of physics such as electromagnetic theory.

\subsection*{Acknowledgment}

Support of the Natural Science and Engineering Research Council of Canada is
gratefully acknowledged.

\bibliographystyle{phreport}
\bibliography{bay,geomalg,phystxts}

\end{subequations}
\end{document}